\documentclass[twocolumn]{aastex631}
\usepackage{amsmath}
\usepackage{savesym}
\savesymbol{tablenum}
\usepackage{siunitx}
\restoresymbol{SIX}{tablenum}
\usepackage{verbatimbox}
\usepackage{soul}
\usepackage{xcolor}
\usepackage{enumitem}
\usepackage{float}

\newcommand{\oiii}{[O\;III]\;$\lambda\lambda$4959,5007}
\newcommand{\ot}{[O\,III]}
\newcommand{\ofive}{[O\;III]\,$\lambda$5007\ }
\newcommand{\ha}{H$\alpha$}
\newcommand{\nii}{[N\,II]}
\newcommand{\firstm}{$\Delta$V$_\mathrm{[O\;III]}$}
\newcommand{\secondm}{$\sigma_\mathrm{[O\;III]}$}
\newcommand{\loiii}{L$_\mathrm{\ot}$}
\newcommand{\lbol}{L$_\mathrm{bol,15\mu m}$}
\newcommand{\sbunit}{erg s$^{-1}$ cm$^{-2}$ arcsec$^{-2}$}

\shorttitle{The outflow impacts on the size of the narrow-line region among type-2 AGNs}
\shortauthors{Jensangjun et al.}
\graphicspath{{./}{}}

\begin{document}

\title{The outflow impacts on the size of the narrow-line region among type-2 AGNs}

\author[0009-0003-8182-1539]{Kantapon Jensangjun}
\author[0000-0002-0269-0135]{Suraphong Yuma}
\email{suraphong.yum@mahidol.edu}
\affiliation{Department of Physics, Faculty of Science, Mahidol University, Bangkok 10400, Thailand}
\begin{abstract}

We present a study of the gas kinematics in narrow-line regions (NLRs) of $2{,}009$ type-2 AGNs at $z<0.34$. 
We construct the \ofive\ emission-line images using publicly available broadband images from the Sloan Digital Sky Survey (SDSS). 
The \ot\ emission area of the samples, measured down to the rest-frame isophote of $1.4\times10^{-15}$ \sbunit, ranges from $3.7$ kpc$^2$ up to $224$ kpc$^2$. 
With our broadband technique, we found a correlation between \ot\ area and AGN luminosity inferred from the \ot\ luminosity and the mid-infrared luminosity at the rest-frame $15\mu$m. 
The isophotal threshold used to determine the \ot\ area affects the correlation strength in that the brighter isophote yields the stronger correlation between the \ot\ area and AGN luminosity. 
The presence of gas outflow is examined by the ratio of the \ot\ velocity dispersion to the stellar velocity dispersion ($\sigma_{\rm \ot}/\sigma_\star > 1.4$) using the SDSS spectra. 
At the given luminosity, the objects with and without outflows exhibit the same extension of the \ot\ emission. 
Their correlation between the \ot\ area and luminosity is almost identical. 
It is suggested that the size of NLRs is not affected by outflow mechanisms but rather by photoionization from the central AGNs. 

\end{abstract}

\section{INTRODUCTION} \label{sec:intro}

Observational evidence on the co-evolution between galaxy properties and the growth of supermassive black holes (SMBHs) at the galactic center has increased significantly over the past few decades \citep[e.g.,][]{Silk&Ree1998,Gultekin2009,Kormendy2013,Baldassare_2020}.
The evolution of the BH accretion rate and the star formation rate density are coincidentally found to be similar; they reach a peak at $z\sim2$ and decrease until the present epoch \citep[e.g., ][]{Madau1996, Hopkins2007, Aird2015, Carraro_2020}, supporting the idea that the evolution of SMBHs and galaxies is closely connected \citep{Lynden1969,Osterbrock2006}.

The hydrodynamic simulations by \cite{Hopkins_2010} show that BH accretion and the star formation rate (SFR) depend on the availability of the cold gas. When the gas is used up, the star formation terminates and the luminosity of the active galactic nucleus (AGN) fades rapidly, marking the galaxy as the red elliptical galaxy with little-to-no SFR or BH accretion \citep{Lapi_2006}.
This coupling has been proposed to explain the observed correlation between the mass of SMBH and the bulge stellar mass \citep{Magorrian_1998, Scott_2013}, or the bulge stellar velocity dispersion \citep{Kormendy2013, Saglia_2016}.

It is believed that the energy released from AGN might play a crucial role in regulating the gas cycle in the host galaxy known as AGN feedback \citep{Croton2006, Somerville_2008, Fabian_2012}.
Current cosmological models of galaxy evolution \citep[e.g.,][]{Bower2006,Croton2006,Somerville2015,Pillepich2019} require AGN feedback to suppress star formation in massive galaxies to reproduce the observed galaxy luminosity function at the high-luminosity end. 
The energy emitted from the central AGNs disturbs the cold gas reservoir by either heating (radiative feedback) or expelling them as outflows (mechanical feedback).
The gas outflows have been detected by numerous studies \citep[e.g.,][]{Woo2016,Torres-Papaqui_2020}, however, their impact on the global SFR of the host galaxy evolution is still unclear. 
Some studies have found that outflows can deplete the cold gas in the interstellar medium (ISM), which is used to form new stars \citep{Wylezalek2016,Chen2022}, terminating the star formation.
Conversely, outflows are found to enhance the SFR by triggering the formation of new stars within outflows themselves, or by feeding the gas back to the interstellar medium  \citep{Shin2019,Zhuang2020,Riffel_2024}. 
Verifying the outflow impacts on the AGN and host galaxies is difficult as it requires knowledge of various parameters such as spatial extent of an outflow, outflow velocity, mass loss rate, etc. 

One approach to study gas outflows driven by AGNs is to investigate the most extensive part of the AGNs called the narrow-line region (NLR). 
It has a size of $0.1-10$ kpc extended from the nucleus, which is spatially resolved in nearby galaxies.
The NLR consists of gas clouds ionized by the central AGN, exhibiting multiple emission lines (e.g., \oiii, [S II]$\lambda\lambda6716, 6731$, [O II]$\lambda\lambda 3729, 3726$).
\citet{Joh2021} suggested that gas clouds in the NLR are transferred from the central AGNs by outflows as they exhibit higher velocity dispersion compared to the gas in star-forming galaxies at fixed stellar mass.
Based on their findings, AGNs with outflows would have larger NLRs than those without outflows, since the gas cloud can propagate farther due to radiation pressure from outflows than by just photoionization from AGNs.
The size of the NLR is usually measured by the \ofive emission lines (hereafter, \ot), as it is one of the most prominent emission lines. The \ot\ line has also been extensively used to study the warm ionized outflows \citep{Woo2016,Kang2018}. 
Hence, using this emission line could provide insight into both the spatial extent and the outflow kinematics of the NLR.

\citet{Cardamone2009} identified galaxies with extremely strong \oiii\ emission lines using the magnitude difference between two broadband images with and without contributions of the emission line. This technique has been used over the past decade by several studies \citep[e.g., ][]{Izotov_2011, Liu_2022}. 
Using similar technique but with the narrowband images, various studies were able to classify galaxies with spatially extended emission lines and related them to outflow mechanisms \citep[e.g.,][]{Yuma_2013,Harikane_2014,Yuma2017,Yuma_2019}. 
Recently, \citet{Sun_2018} used the broadband images from the Subaru Hyper Suprime-Cam (HSC) survey to detect the resolved \ot\ emission in a sample of $300$ type-2 AGNs.
They have found a strong correlation between the \ot\ area and AGN luminosity. 
However, the correlation of \ot\ emission with gas kinematics or outflows has not been yet explored.

In this work, we aim to investigate the impact of outflows on the size of NLR. 
We constructed the \ot\ images using the broadband images and spectra from the Sloan Digital Sky Surveys \citep[SDSS,][]{York2000}. 
The gaseous kinematics and outflows are determined from the \ot\ emission line taken from the SDSS spectroscopic data in the integrated $\SI{3}{\arcsecond}$ fiber.
The paper is organized as follows. Section \ref{sec:data} provides information on our samples and data.
The construction of \ot\ images and the area measurement are explained in Section \ref{section:oiiiimg}. 
We described the \ot\ kinematics and outflow selection in Section \ref{sec:oiiikinematics}.
Our results are shown in Section \ref{sec:results}.
We discuss the results of the size-luminosity relations and the outflow impacts in Section \ref{sec:discuss} and summarize in Section \ref{sec:conclusion}. 
Throughout this work, we assumed a flat $\Lambda$CDM cosmology with an $h=0.7$, $\Omega_m=0.3$ and $\Omega_\Lambda=0.7$.

\section{SAMPLE AND DATA}\label{sec:data}
\subsection{Sample Selection}

\begin{figure}[ht!]
\epsscale{1.2}
\centering
\includegraphics[width=80mm]{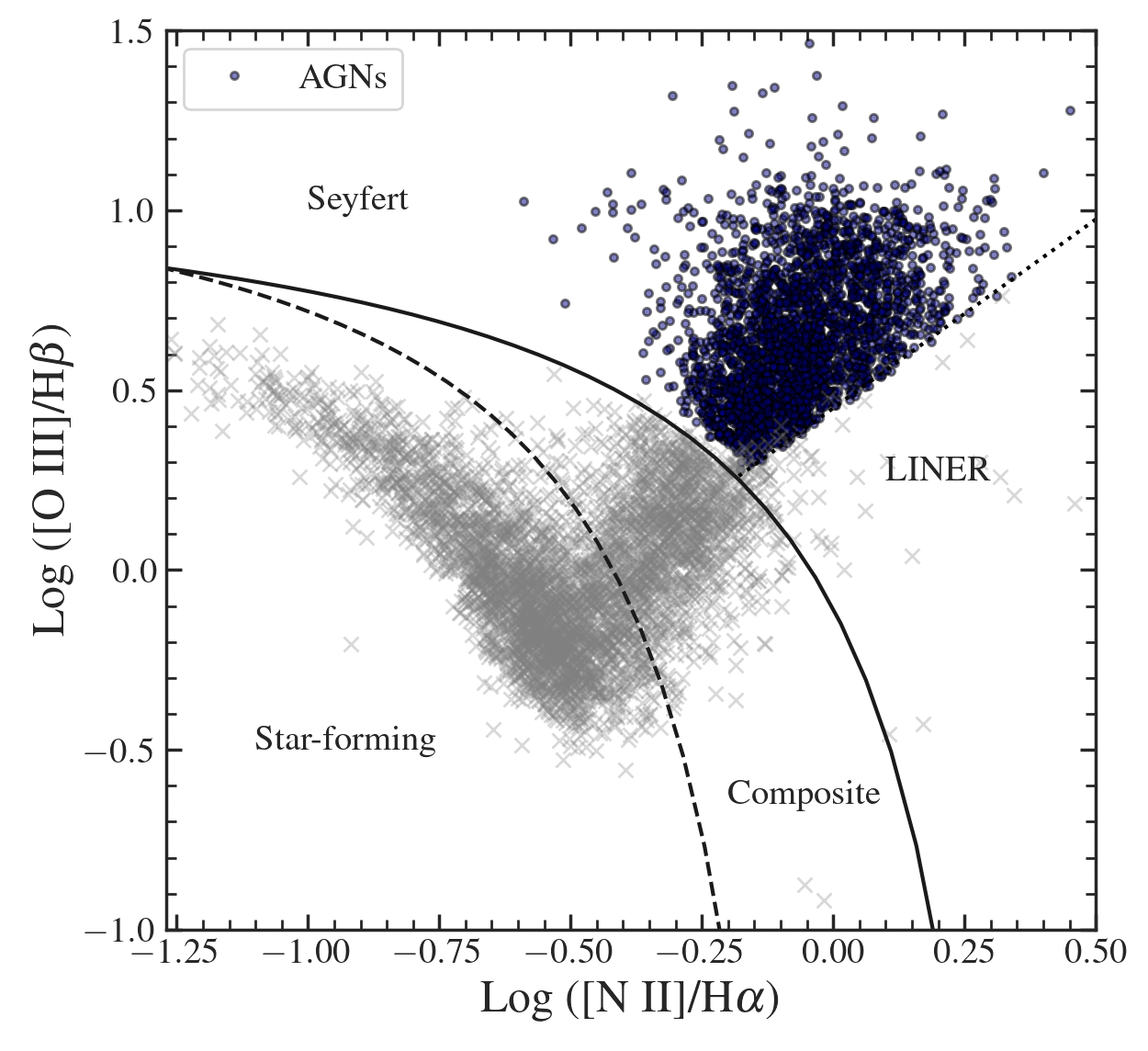}
\caption{The [N II]-based BPT diagnostic diagram of $6{,}673$ in the ``emissionlinesport" table. Navy blue dots denote our parent sample of $2{,}724$ type-2 AGNs. Gray crosses represent galaxies classified as non-AGNs. The dashed line shows the empirical division between star-forming galaxies (SFs) and AGNs from \citet{Kauffmann2003}. The solid line represent the theoretical demarcation line between SFs and AGNs presented by \citet{Kewley_2001}. The dotted line shows the separation between Seyfert and LINER from \citet{Schawinski_2007}.} \label{fig:bpt}
\end{figure}

\begin{figure*}[ht!]
\epsscale{1.2}
\centering
\includegraphics[width=180mm]{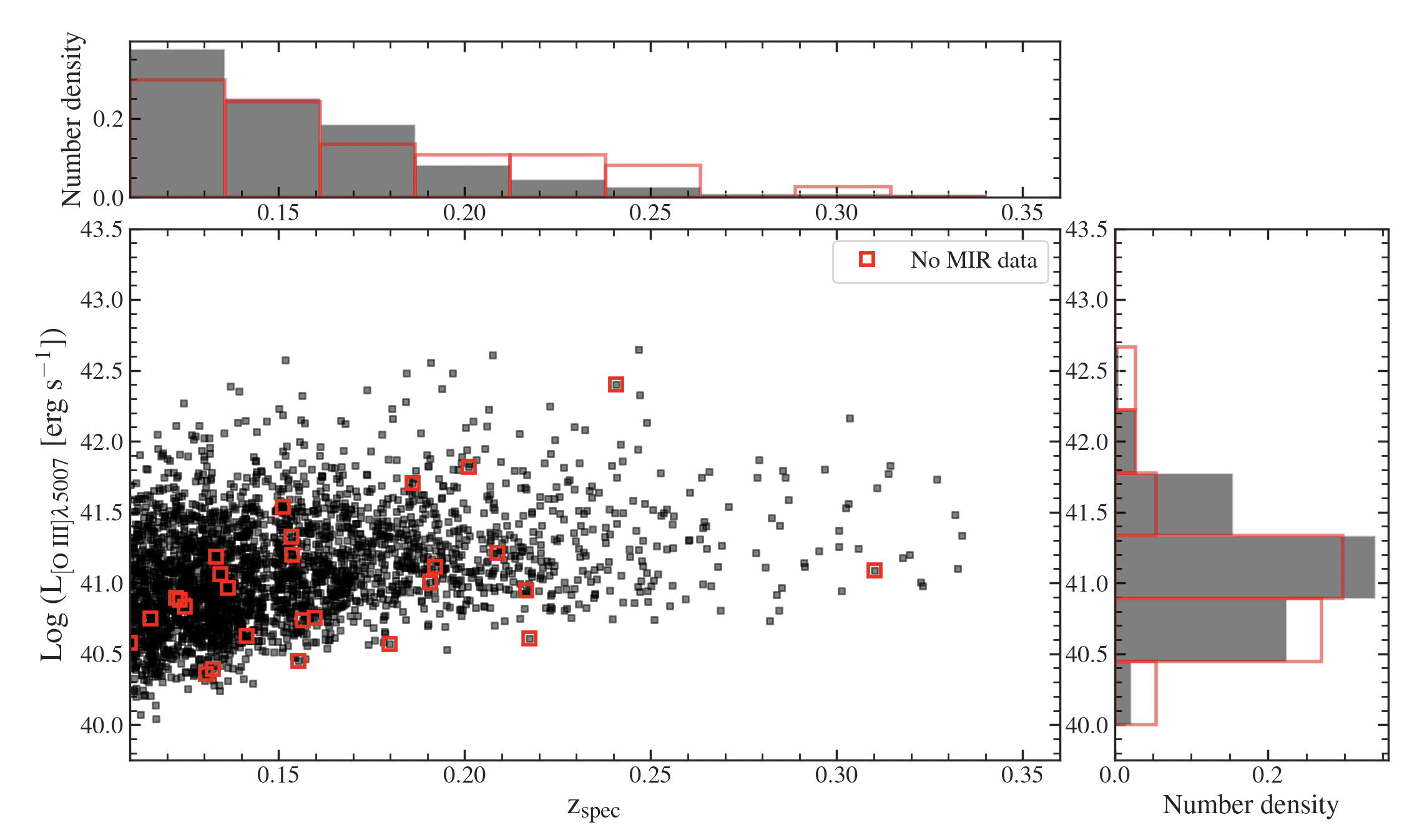}
\caption{L$_\mathrm{[O\;III]\lambda5007}$ vs spectroscopic redshift ($z_\mathrm{spec}$) of $2{,}724$ type-2 AGNs at $0.13< z < 0.34$. The solid black and open red squares represent the samples with and without WISE/MIR detection, respectively (see Section \ref{sec:wise}). The histograms in the x-axis and y-axis show the distribution of $z_\mathrm{spec}$ and L$_\mathrm{[O\;III]\lambda5007}$ for samples with MIR detections (black) and no MIR detections (red), respectively.} \label{fig:Loiii-z}
\end{figure*}

A sample of type-2 AGNs at $0.13< z < 0.34$ was selected from objects classified as ``Seyfert" in the SDSS ``emissionlinesport" table reported by \cite{Thomas2013}. 
The ``Seyfert" label in the catalog was assigned based on the [N II]-based BPT diagnostic diagrams \citep{ Baldwin_1981, Kewley_2001, Schawinski_2007}, which plot the emission-line ratios of [O III] $\lambda5007$/H$\beta$ and [N II] $\lambda6583$/H$\alpha$. 
The redshift ranges are calculated such that the \oiii\ emission lines fall within the full width at half maximum (FWHM) of the SDSS-$r$ filter.
Figure \ref{fig:bpt} illustrates the [N II]-based BPT diagram of all $6{,}673$ galaxies in the ``emissionlinesport" table with S/N $>5$.
We finally obtained a sample of $2{,}724$ type-2 AGNs with \ot\ luminosity in the range of $\log(L_\mathrm{[O\;III]}/\mathrm{erg\,s^{-1}}) = 40.0-43.0$. 
In Figure \ref{fig:bpt}, we see that all selected samples have [O\;III] emission line dominated over the H$\beta$ emission line with a minimum ratio of 1.97, suggesting that the contribution of the [O\;III] emission line is stronger than H$\beta$ in the emission-line images.
Figure \ref{fig:Loiii-z} shows the redshifts and the dust-corrected \ot\ luminosity of our samples.
The \ot\ luminosity in our samples is evenly distributed over the redshift range. 

\subsection{SDSS Images and Spectra}

The broadband images and spectra are retrieved from SDSS Data Release 12 \citep{Alam2015}. We used broadband $r, i$ and $z$ images covering the wavelength range of $5{,}400$\AA\ $-11{,}000$\AA\ with the median $5\sigma$ limiting magnitude for point sources of $22.70, 22.20$, and $20.71$ mag, respectively.
All the images are registered to the World Coordinate System (WCS) with a pixel scale of \SI{0.396}{\arcsec}, 
which corresponds to $0.8-1.9$ kpc at the redshift range of our sample. For each object, we cropped the image to a size of \SI{40}{\arcsec}$\times$\SI{40}{\arcsec} centering at the target's position in the $r$-band image.

The spectral data are taken with the SDSS spectrograph with a fiber diameter of \SI{3}{\arcsec} corresponding to a physical diameter of $6.0$--$14.5$ kpc. The wavelength coverage is $3{,}800$--$9{,}200$ \AA\ with a resolution of $R\sim 1{,}500$ at $3{,}800$ \AA, and $R\sim 2{,}500$ at $9{,}000$ \AA. 

\subsection{Mid-infrared data from WISE}\label{sec:wise}

The mid-infrared (MIR) data were obtained from the Wide-field Infrared Survey Explorer \citep[WISE;][]{WISE2010}. The survey has been conducted in four bands: W1 (3.4$\mu$m), W2 (4.6$\mu$m), W3 (12$\mu$m), and W4 (22$\mu$m). The positions and photometry of the WISE/MIR objects are described in the WISE ALL-Sky release source catalog\footnote{\url{https://wise2.ipac.caltech.edu/docs/release/allsky/expsup/sec2_2a.html}}. 
We adopted the ALL-Sky catalog over the more recent ALLWISE catalog because the number of objects detected in the W3 and W4 bands is higher in the ALL-Sky catalog.
We cross-matched our AGN sample with the WISE/MIR objects within a radius of $1\arcsec$. We adopted the same matching radius as used by \citet{Greenwell2023}, which covers $86.4\%$ of the closest SDSS-WISE associates. As a result, we obtained 2{,}634 ($\sim97\%$) MIR counterparts, while $130$ type-2 AGNs in our sample had no match in the WISE ALL-Sky catalog (open red squares in Figure \ref{fig:Loiii-z}).

We estimate the bolometric luminosity by using the mid-infrared luminosity to instead of generally used \loiii\ \citep{Heckman2004}, as the \ot\ emission can be affected by uncertainties from the dust correction model. The mid-infrared emission is less sensitive to dust absorption than the optical emission lines. The detailed calculation of the bolometric luminosity is explained in Section \ref{sec:area-lum}. 

\section{\ofive spatial extension}\label{section:oiiiimg}

We determined the \ofive spatial extension by constructing the \ot\ emission-line images from the SDSS broadband images. 
In short, the $r-$band images containing the \ot\ emission were subtracted by the stellar continuum estimated at the same effective wavelength. Then the continuum subtracted images were rescaled to correct for the \ofive contribution to the $r$ band. 
The detailed procedures are explained below. 

\subsection{Continuum images}\label{sec:interpolation}

\begin{figure*}[ht!]
\epsscale{1.2}
\centering
\includegraphics[width=180mm]{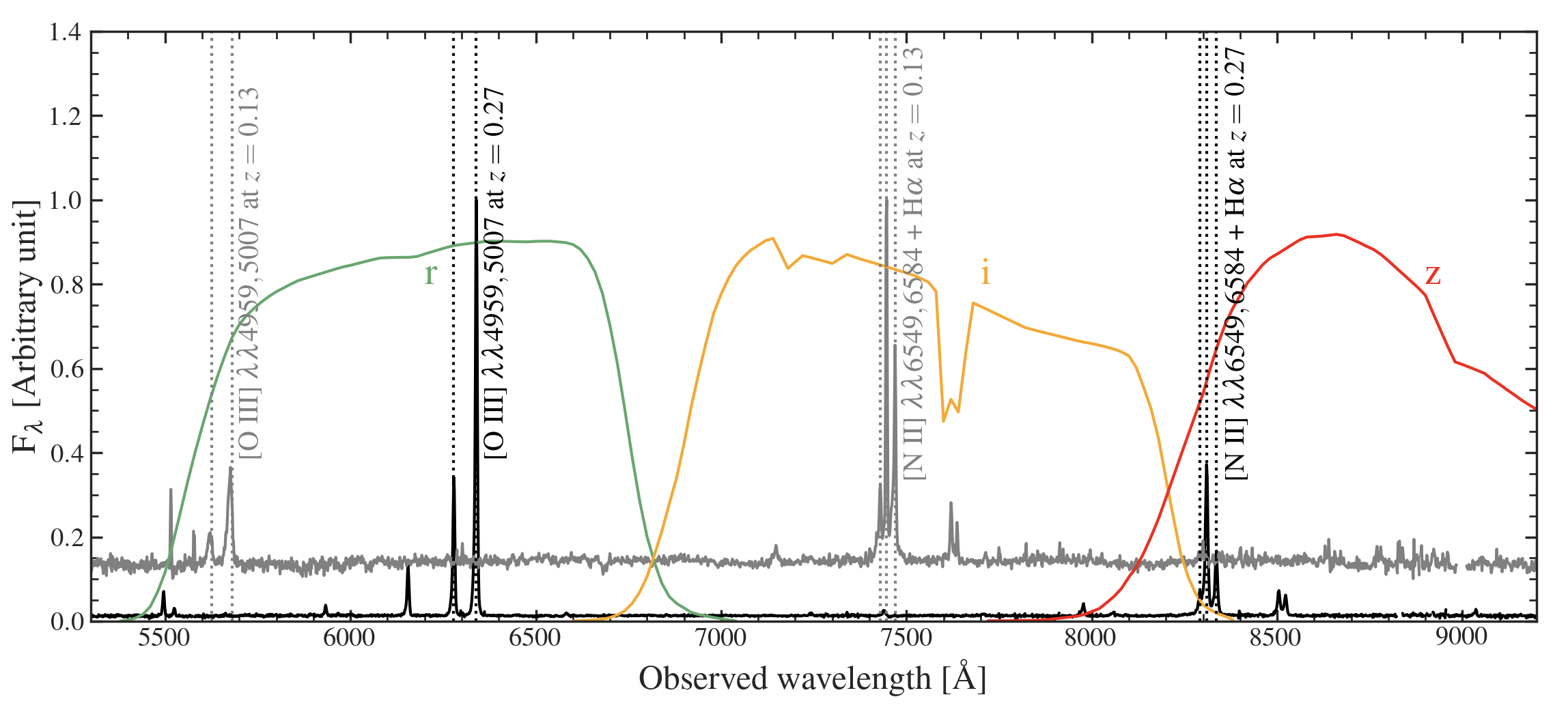}
\caption{
Spectra of SDSS J$092729.11+193640.1$ at $z=0.13$ (gray) and SDSS J$162209.41+352107.5$ at $z=0.27$ (black). 
The green, yellow, and red lines represent the area normalized transmission curve of $r$, $i$, and $z$ bands, respectively. 
The vertical dashed lines indicate the location of emission lines. 
}
\label{fig:specwithfilters}
\end{figure*}

The stellar continuum in the $r$-band image was estimated from another broadband image that is not contaminated by strong emission lines such as [N II]$\lambda\lambda 6548, 6584$ and H$\alpha$. 
For $2{,}609$ objects at $0.13<z<0.26$, the continuum image was constructed from $z$ images, while the $i-$band image was used for the remaining $115$ samples at $0.26<z<0.34$ where the [N II]+H$\alpha$ shifts into the $z$ band. 
Figure \ref{fig:specwithfilters} shows spectra of two examples at both redshift ranges. 
The \nii\ and \ha\ emission of SDSS J$092729.11+193640.1$ at $z=0.13$ fall into the $i$-band filter, so we estimated its stellar continuum from the $z$ band. Likewise, SDSS J$162209.41+352107.5$ at $z=0.27$ has both emission lines in the $z$ band. The $i-$band image was used to determine the continuum instead. 

Because the transmission curves and bandwidths are different in all $riz$ bands, the $z$-band and $i$-band images are scaled to match the continuum in the $r$-band image. 
Generally, the ratio of the stellar continuum in the $r$ band to that in the $i$ or $z$ band can be determined by convolving the spectrum with the transmission curves of the corresponding filters. 
However, the wavelength coverage of the spectra is up to 9200 \AA, while it is 11,000 \AA\ for the $z$ band. 
Instead of using the observed spectra, We fit them with the power law 
and use the best-fit power-law spectrum to convolve with the filters. 
The ratio ($\Theta$) of the continuum in the $r$ band to that in the $i$- or $z$-band is thus written as

\begin{equation}
     \Theta = \frac{\int \lambda^\beta T_r \lambda\;d\lambda}{\int \frac{T_{r}}{\lambda}d\lambda} \frac{\int \frac{T_{i,z}}{\lambda}d\lambda}{\int \lambda^\beta T_{i,z} \lambda\;d\lambda}
     \label{eq:extinctfactor}
 \end{equation}

where $T_r$ and $T_{i,z}$ are the transmission curves of the $r$ and $i$ or $z$ bands, respectively, and $\beta$ is the best-fit spectral index.

\subsection{The point spread function (PSF) matching}\label{sec:psfmatching}

\begin{figure*}[ht!]
\centering
\includegraphics[width=160mm]{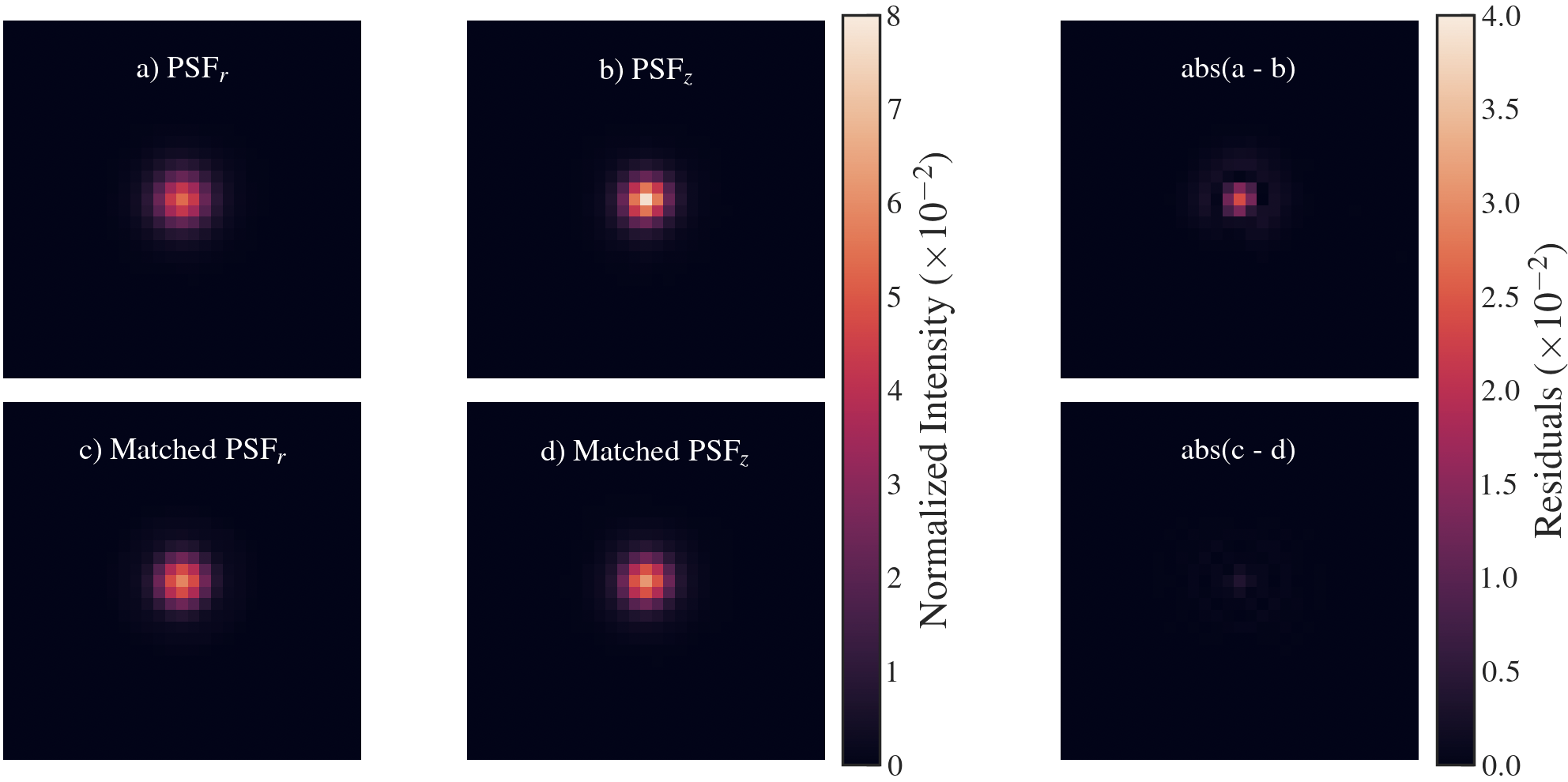}
\epsscale{1.2}
\caption{
Example of the PSF models of the $r$-band image (left), the $z$-band image (middle), and the residual (right). The colorbar represents the normalized intensity such that the sum equals unity. The top and bottom panels show the PSF models before and after the matching using the Gaussian kernel, respectively. 
}
\label{fig:psfmatching}
\end{figure*}

We reconstructed the PSF model of each image using the psField files from the SDSS Science Archive Server (SAS)\footnote{See PSF: \url{https://www.sdss4.org/dr17/imaging/images/\#psf}}. 
In the case of the continuum images, we adopted the PSF model in the filter originally used to create the continuum image (i.e., $i-$ or $z-$ band). We homogenized the PSF size of both $r-$band and continuum images 
using the $2$-dimensional Gaussian kernel to obtain the target FWHM of $1.93$\arcsec, the maximum PSF size of all images. 
Figure \ref{fig:psfmatching} shows an example of the PSF matching between $r-$band and $z-$band images. 
The original FWHM of the examples $r-$band and $z-$band images are $1.71\arcsec$ and $1.64\arcsec$, respectively. 
The top panel illustrates that the images can be oversubtracted without the PSF matching. 
In this study, we keep the maximum residual below 10\% of the peak of the target PSF by adjusting the width of the Gaussian kernel.

\subsection{Construction of [O III] images}\label{sec:oiiiconstruction}

After the PSF matching, we align images from different bands according to their WCS headers. The continuum images are adjusted to match the position in the $r$ band and then subtracted from the $r-$band image. 
The emission-line image contains all emission lines falling into the $r$ band. Although the \ot\ emission line is normally the strongest, the contribution of the other lines in the emission-line image could cause an overestimation in \ot\ area measurement. Using the spectrum of each object, we calculated the ratio of the \ot\ emission line to the total fluxes from all emission lines in the $r$ band (hereafter called $\gamma_{5007}$). The $\gamma_{5007}$ was used to scale the emission-line images to the \ot\ emission images.
We also corrected the effects of the cosmological dimming and transmission curves as the objects are slightly at different redshifts and their \ot\ emission falls into different parts of the $r$-band filter.
Ultimately, the intensity of the \ofive image ($I_\mathrm{[O\;III]}$) can be calculated from the intensity of the emission-line image ($I_{em}$) as follows   

\begin{equation}
I_\mathrm{[O\;III]} =  \frac{\gamma_{5007}\times(1+z)^4\times c}{T_{r,5007}\times\lambda_{\mathrm{obs},5007}} \times {\int \frac{T_{r}}{\lambda}d\lambda} \times I_{em}
\label{eq:icon}
\end{equation}
 where $\gamma_{5007}$ and $\lambda_{\mathrm{obs},5007}$ are the flux ratio of \ofive to total in the $r$ band and the observed wavelength of \ofive. $T_{r,5007}$ and $T_r$ are the transmission function of the $r$-band filter at $\lambda_{\mathrm{obs},5007}$ and any wavelength $\lambda$, respectively. $c$ is the speed of light in vacuum.

\subsection{Isophotal area of [O III] $\lambda5007$ emission}\label{sec:oiiiarea}

\begin{figure}[ht!]\epsscale{1.4}
\includegraphics[width=85mm]{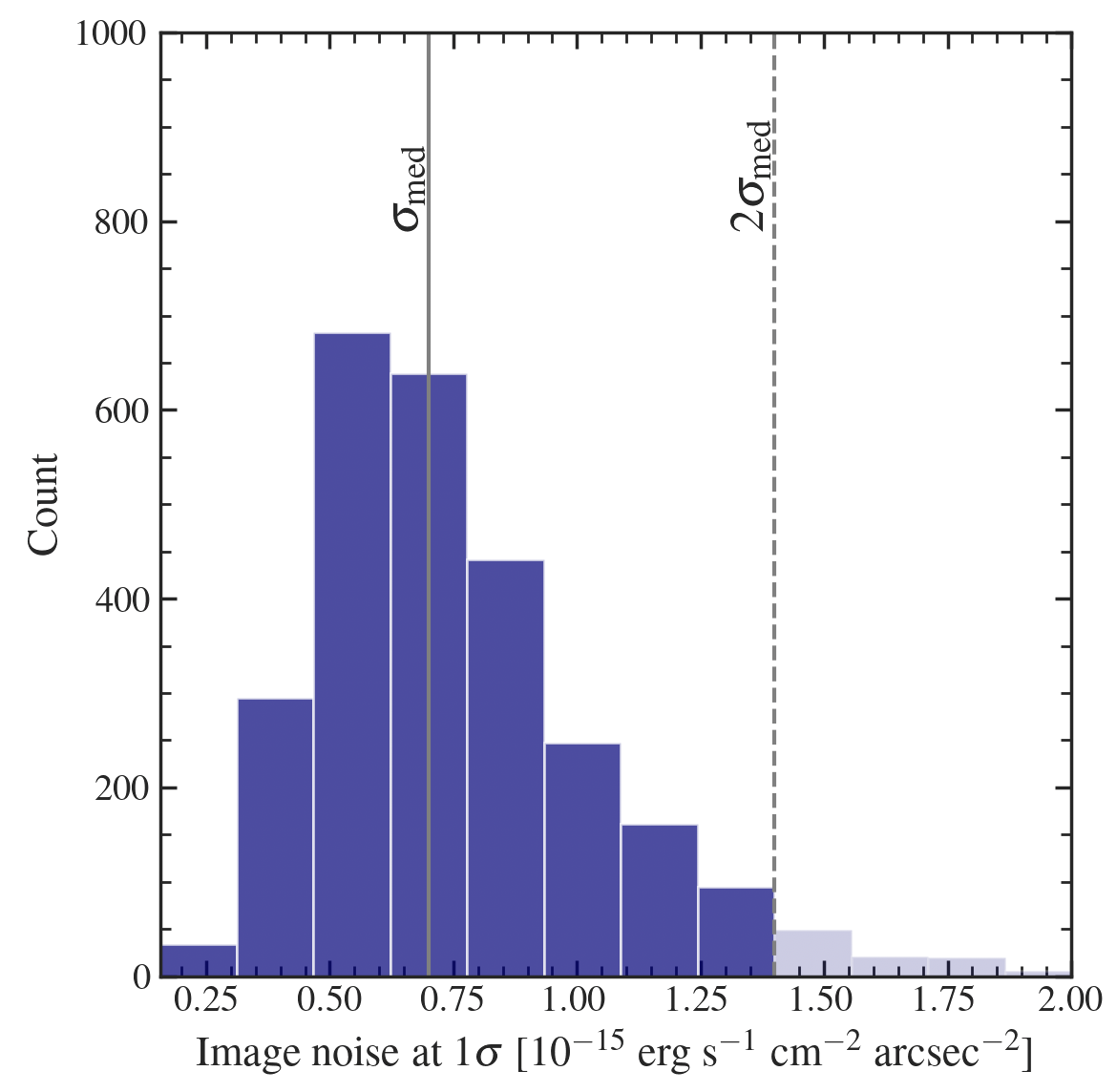}
\caption{The noise distribution of $2{,}724$ [O\;III] emission line images of $2{,}724$  samples. The median noise $\sigma_\mathrm{med}$ of $7.0\times10^{-16}$ \sbunit\ is illustrated by the solid line, while the dashed line shows $2\sigma_\mathrm{med}$ level. The shaded histograms show $127$ samples with $\sigma>2\sigma_\mathrm{med}$, which are removed from further discussion. \label{fig:noise-EM}}
\end{figure}

We measured the depth of [O\;III] images by fitting the noise distribution of each emission-line image. A histogram of $1\sigma$ noise level of all [O$\;$III] images is shown in Figure \ref{fig:noise-EM}. The median of 1$\sigma$ noise ($\sigma_\mathrm{med}$) is $7.0\times10^{-16}$ \sbunit. We adopted $2\sigma_\mathrm{med}$ ($1.4\times10^{-15}$ \sbunit) as a flux limit in measuring the isophotal area of the target in the \ot\ images. To avoid noise contamination in the area measurement, we excluded $127$ objects with the \ot\ images showing $1\sigma$ noise levels higher than $2\sigma_\mathrm{med}$ marked by the dashed line in Figure \ref{fig:noise-EM}. 
This isophotal threshold goes shallower than the threshold of $3\times10^{-15}$\sbunit\ adopted by \citet{Sun_2018} using broadband images from the Subaru/HSC. However, it is still deeper than the typically used value of $1\times10^{-15}$ \sbunit\ adopted in other studies \citep{Liu2013,Hainline2013} using IFU/long-slit spectroscopy to detect [O\;III] images.

To avoid possible emission from nearby galaxies or foreground stars, we only considered the emission whose center is located within $5^{\prime\prime}$ ($11.6-24.2$ kpc) from the center of the stellar continuum. 
Notably, this method possibly excludes flickering AGNs that do not show central emission but extended light echo at a larger distance due to the past ionization \citep{schawinski2015}. 

The uncertainty in the \ot\ area was determined by a set of one hundred Monte Carlo simulations. We randomly added Gaussian noise with 1$\sigma$ fluctuation into the \ot\ image. For objects with $\sigma<\sigma_\mathrm{med}=7.0\times10^{-16}$ \sbunit\, we alternately inserted the noise of $\sigma_\mathrm{med}$. The \ot\ area of noisy images are measured the same way as the original \ot\ images. The standard deviation of the noise-added images is adopted as the uncertainty of the area denoted as $\sigma_\mathrm{A}$.

\subsection{Accuracy in the \ot\ area measurement}
\begin{figure}[ht!]
\epsscale{1.2}
\centering
\includegraphics[width=80mm]{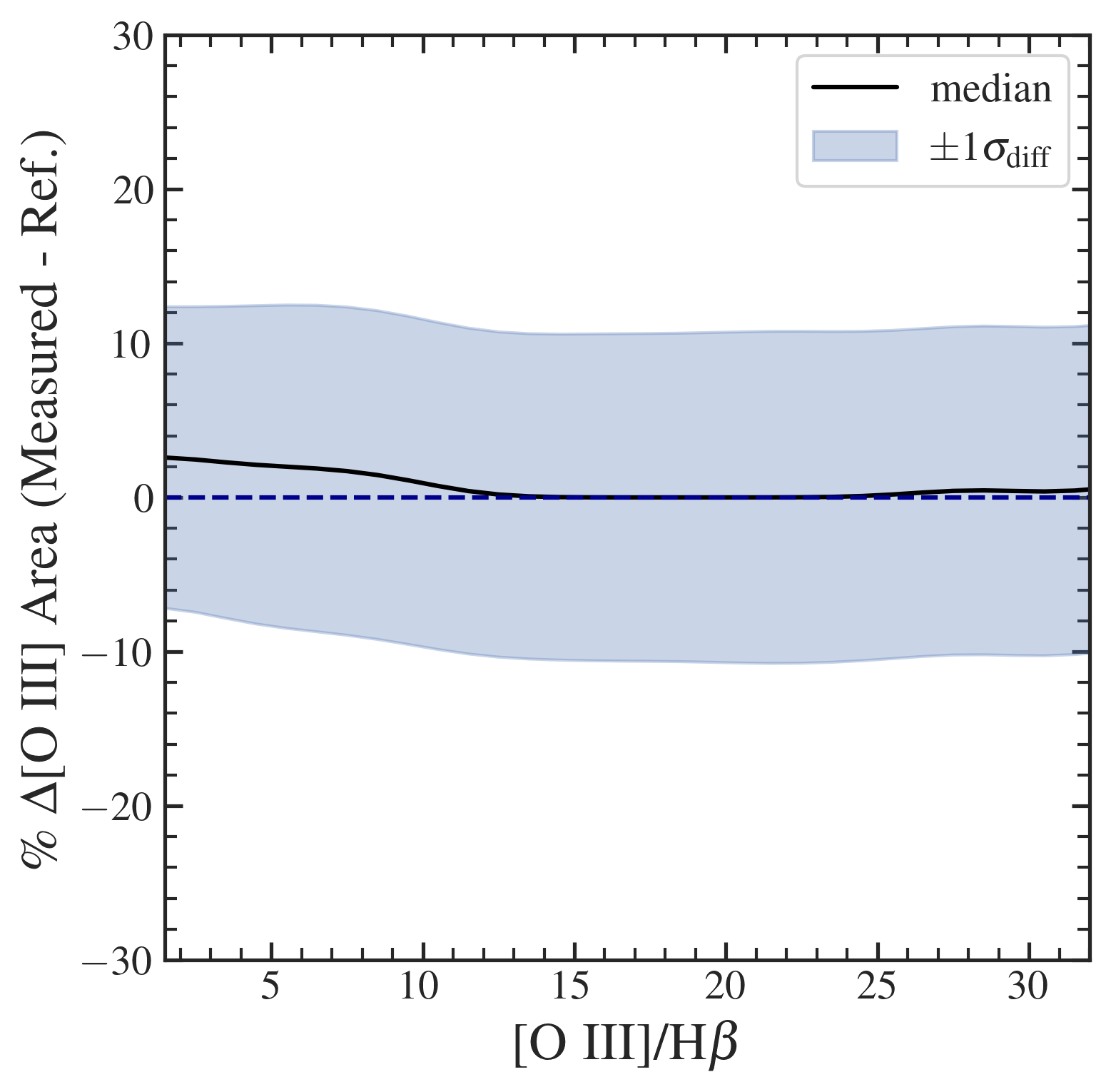}
\caption{The percentage of area difference measured from the \ot+H$\beta$ map and from the reference \ot\ map from MaNGA at each \ot/H$\beta$ ratio. Solid line indicates the median percentage of the isophotal areas from the \ot\ maps, and the blue contour represents the standard deviation ($1\sigma$).}

\label{fig:manga_dif}
\end{figure}

We examine the contribution of H$\beta$ emission line in the constructed emission-line image by simulating SDSS $r$-band images that contain both \ot\ and H$\beta$ emissions at different levels of \ot/H$\beta$ ratios, constructing the \ot\ emission-line image, and recovering the \ot\ area using the same processes as our samples. The simulated images are constructed from the emission-line maps from one of our samples (SDSS J$075221.86+341935.5$, $z\sim0.14$), which has the IFU data in the Mapping Nearby Galaxies at APO (MaNGA). This object has the \ot/H$\beta$ ratio of $2.61$.
Based on the emission-line map of this sample, we created the \ot+H$\beta$ images at different \ot/H$\beta$ ratios ranging from $1.5-40$. The SDSS background is also added into the maps. The contribution of H$\beta$ was then corrected based on the same procedures as our sample to construct the \ot\ emission-line image.  
We performed a set of five hundred Monte Carlo simulations for each \ot/H$\beta$ ratio by varying the inserted background. 
The \ot\ area was measured down to $2\sigma_{\rm med}$ and compared with the reference \ot\ map from MaNGA. 
Figure \ref{fig:manga_dif} shows the percentage difference in the \ot\ isophotal areas between the \ot\ image constructed from the simulated \ot+H$\beta$ map and the reference \ot\ map from MaNGA. The maximum median offset is $\sim2\%$ in cases of \ot/H$\beta$ $< 12$, and the value approaches zero at the higher \ot/H$\beta$ ratio. 
We note that this median offset is negligible and it is within the uncertainty of our area measurement shown in Section \ref{sec:oiii-images}.

\section{Measurement of [O III] kinematics}\label{sec:oiiikinematics}

The gas kinematics in the NLR are examined via the \oiii\ emission lines from the SDSS integrated spectra. We firstly removed the stellar continuum from the spectra. The stellar templates are constructed using the MILES-based models by \citet{vazdekis2010}. We assumed the \citet{Salpeter_1955} initial mass function (IMF), 6 steps of metallicities, [Z/H] = [$-1.71, -1.31, -0.71, -0.40, 0.00, 0.22$], and 26 age steps equally spaced from $1.0$ to $12.6$ Gyrs. We used the penalized Pixel-Fitting package \citep[pPXF,][]{Cappellari2004} to perform the stellar continuum fitting and subtracting.

\begin{figure*}[ht!]
\epsscale{1.2}
\includegraphics[width=180mm]{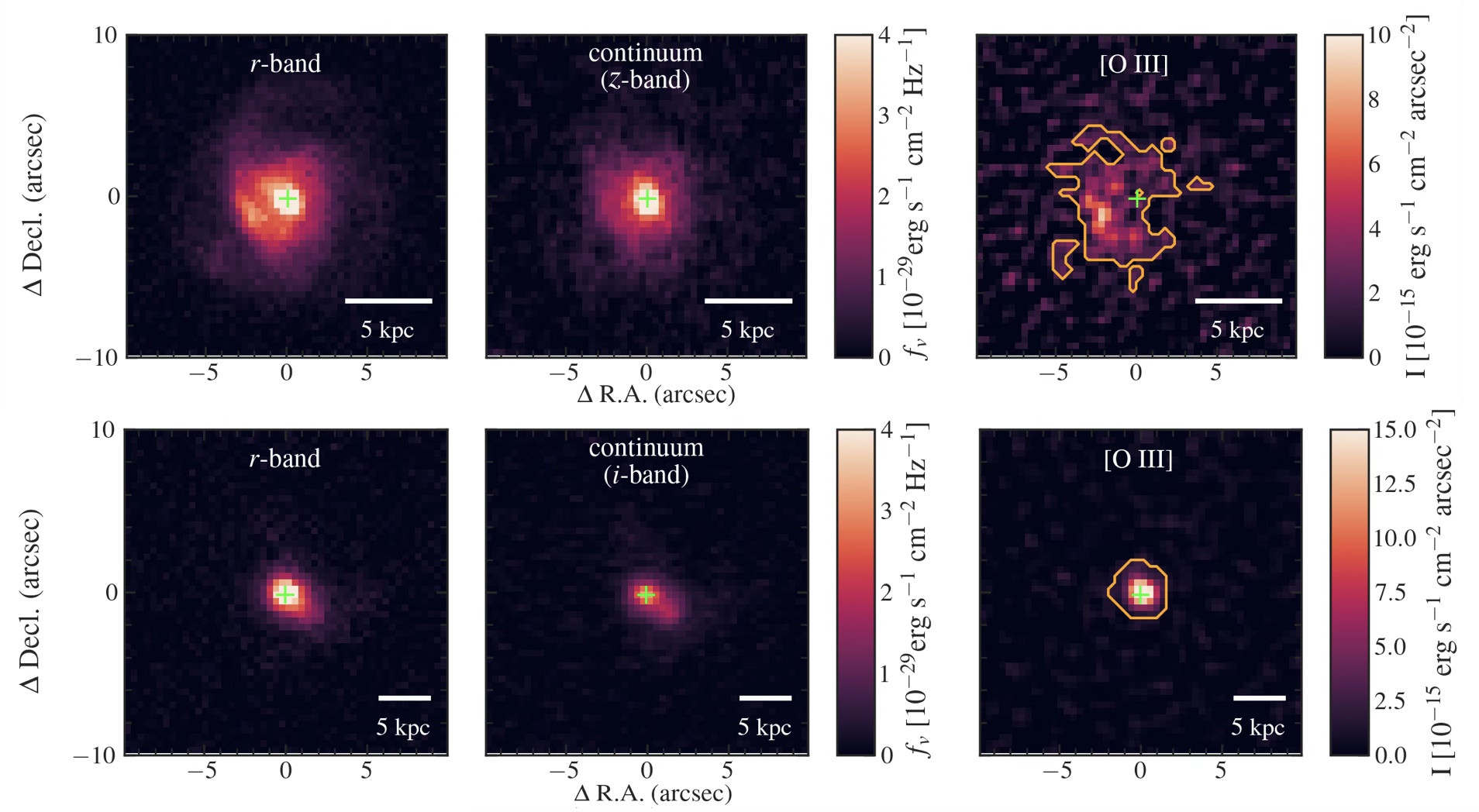}
\caption{\ofive image construction of two examples: SDSS J$092729.11+193640.1$ at $z=0.13$ (top) and SDSS J$162209.41+352107.5$ at $z=0.27$ (bottom). The image size is $20\arcsec\times20\arcsec$. The left, middle, and right panels show the original $r$-band, constructed continuum, and resulting [O\;III] images, respectively. The green crosses mark the center in the continuum image. The colormap in the $r$-band and continuum images represents flux density in a unit of erg s$^{-1}$ cm$^{-2}$ Hz$^{-1}$. The right panel shows the \ot\ image in a unit of $10^{-15}$ erg s$^{-1}$ arcsec$^{-2}$. The contour encloses the isophote area down to $1.4\times10^{-15}$ \sbunit\ (see Section \ref{sec:oiiiarea}). \label{fig:sdss-oiii}}
\end{figure*}

After the stellar continuum subtraction, the \oiii\ emission lines were fitted with Gaussian profiles using the \texttt{MPFIT} package 
\citep{Markwardt2009}, which was translated into \texttt{PYTHON} version by Mark Rivers\footnote{{\texttt{PYTHON} version of \texttt{MPFIT} created by Mark Rivers: }\url{https://cars9.uchicago.edu/software/python/mpfit.html}}. 
The [O\;III] doublets are assumed to share the same kinematics (i.e., identical systemic velocity and line width). 
Each emission line was fitted twice with a single and double Gaussian profile. In the case of single Gaussian fitting, the width of the Gaussian profile must be larger than the spectral resolution of SDSS spectra ($69$ km$\;$s$^{-1}$). In the case of double Gaussian fitting, we constrained the core and wing components to center within $\pm180$ km$\;$s$^{-1}$ and $\pm 500$ km$\;$s$^{-1}$ of the systemic velocity, respectively. The peak of each component must be higher than the noise level, i.e., the amplitude-to-noise ratio (A/N) must be larger than $3$ \citep{Woo2016}. 

We determined the Bayesian information criterion (BIC) for each fit to assess whether the \oiii\ emission lines are better modeled with a single or double Gaussian profile. The model with a lower BIC value is generally chosen over the other. 
We computed $\Delta\mathrm{BIC}$ as $ (\chi^2_\mathrm{1}+k_\mathrm{1}\ln n) - (\chi^2_\mathrm{2}+k_\mathrm{2}\ln n)$, where $\chi^2_\mathrm{1,2}$ are the total $\chi^2$ for the single and double Gaussian profile, respectively, $k_\mathrm{1,2}$ are the number of free parameters of each profile, and $n$ is the number of data points. The double Gaussian profile was adopted for the samples with $\Delta\mathrm{BIC} > 10$ \citep{Swinbank2019}.
 
We determined the \ot\ velocity dispersion (\secondm) from the non-parametric [O\;III] linewidth ($\Delta\lambda$), calculated from the best fitting profile as follows. 

\begin{equation}
    \Delta\lambda = \sqrt{\frac{\int\lambda^2 f_\lambda \,d\lambda}{\int f_\lambda d\lambda} - \lambda_\mathrm{cen}^2}, 
    \label{eq:sigma}
\end{equation}

\noindent where $\lambda_\mathrm{cen}=\int \lambda f_\lambda d\lambda\;/\int f_\lambda d_\lambda$ is the flux-weighted center of the line profile and $c$ is the speed of light in vacuum. 
The \ot\ velocity shift (\firstm) is translated from the shift of $\lambda_\mathrm{cen}$ from the center of the emission line. 

\section{RESULTS}\label{sec:results}

\subsection{Spatial Extent of [O\;III] emission}\label{sec:oiii-images} 

Figure \ref{fig:sdss-oiii} shows the \ot\ image construction of two examples whose spectra are illustrated in Figure \ref{fig:specwithfilters}. 
The top panel is SDSS J$092729.11+193640.1$ at $z=0.13$, whose continuum image was constructed from the $z$-band image, while the continuum image of SDSS J$162209.41+352107.5$ at $z=0.27$ in the bottom panel was created from the $i$-band image. 
The \ot\ image of both objects shows a different morphology than the continuum one.
In addition to 127 objects with noisy \ot\ images, we excluded another 588 samples from the analysis as there is no \ot\ emission in the subtracted images. These omitted objects either have small observed [O III] equivalent width ($<50$ \AA) or their intensity in the z-band images is contaminated by noise. We have $2{,}009$ AGNs to investigate the \ot\ extension.

\begin{figure*}[ht!]
\centering
\includegraphics[width=140mm]{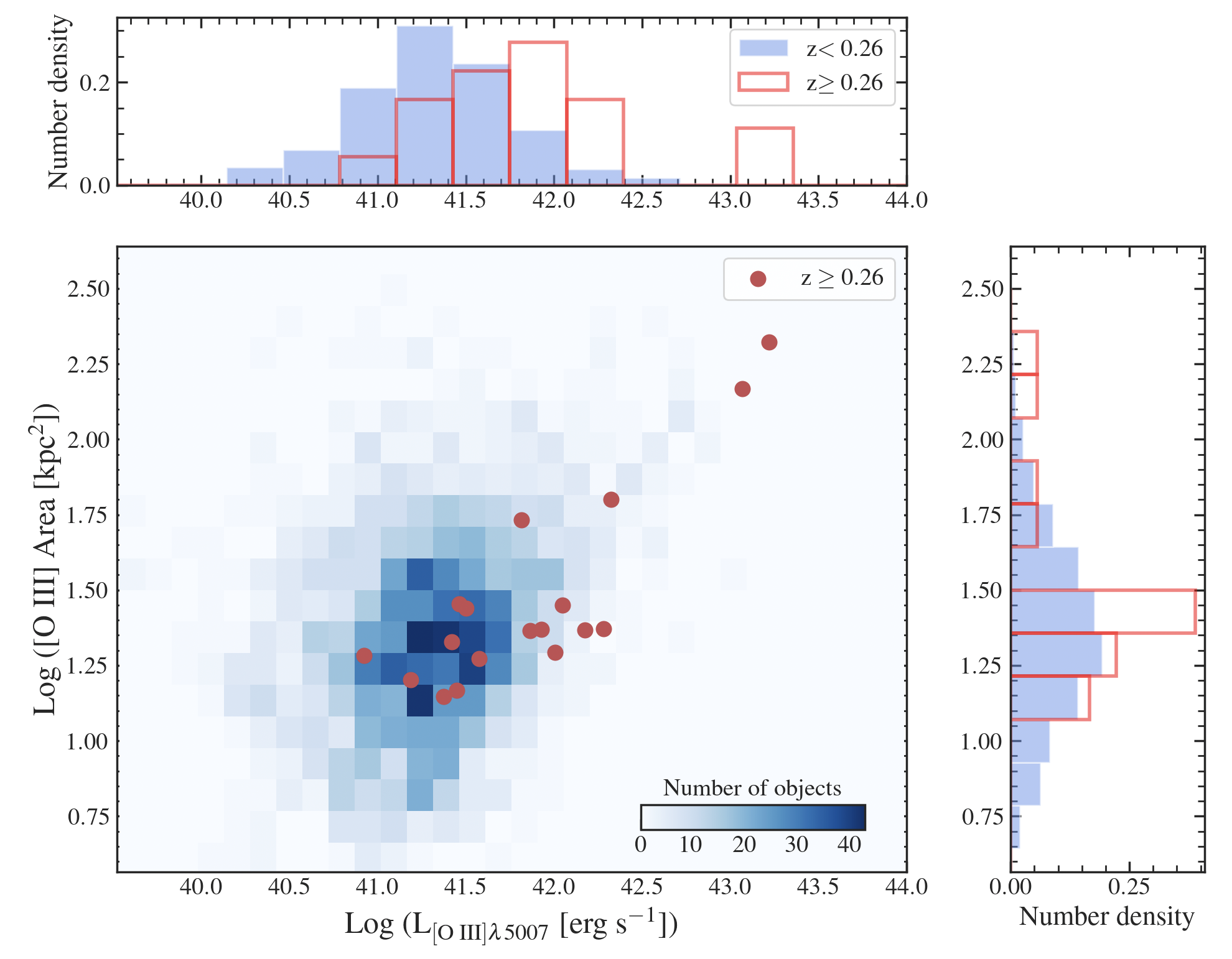}
\epsscale{1.2}
\caption{\ot\ area-luminosity relation of $2{,}009$ AGNs. The density map in the middle panel indicates the number of samples at $z<0.26$. The red dots are samples at $z\geq0.26$. The top and right panels show the distributions of \ot\ area and \ot\ luminosity, respectively. Shaded and open histograms represent samples at $0.13<z<0.26$ and at $z\geq0.26$, respectively.}
\label{fig:areadist}
\end{figure*}

Figure \ref{fig:areadist} shows the overall relation between \ot\ area and \ot\ luminosity of $2{,}009$ AGNs. 
The \ot\ areas range from $3.7$ kpc$^2$ to $224$ kpc$^2$, while the \ot\ luminosity is approximately \loiii$\sim10^{40-43}$ erg s$^{-1}$. 
The \ot\ area tends to increase with increasing \ot\ luminosity, especially for objects at $z\geq0.26$. 
The objects at $z\geq0.26$ are shown separately as their continuum images were constructed using the $i$-band instead of the $z$-band used at the lower redshifts (see Section \ref{sec:interpolation}). At $z\geq0.26$, the samples tend to exhibit larger \ot\ area and higher \loiii\ than those at $z<0.26$. 
Based on the Kolmogorov-Smirnov (KS) test, we can reject the null hypothesis that the distributions of the \ot\ area and luminosity of the objects at $z<0.26$ and at $z\geq0.26$ are drawn from the same populations at the p-value of $3.47\times10^{-6}$ and $0.01$, respectively.

\subsection{[O\;III] area-AGN luminosity relation}\label{sec:area-lum}  

\begin{figure*}[ht!]
\epsscale{1.2}
\plotone{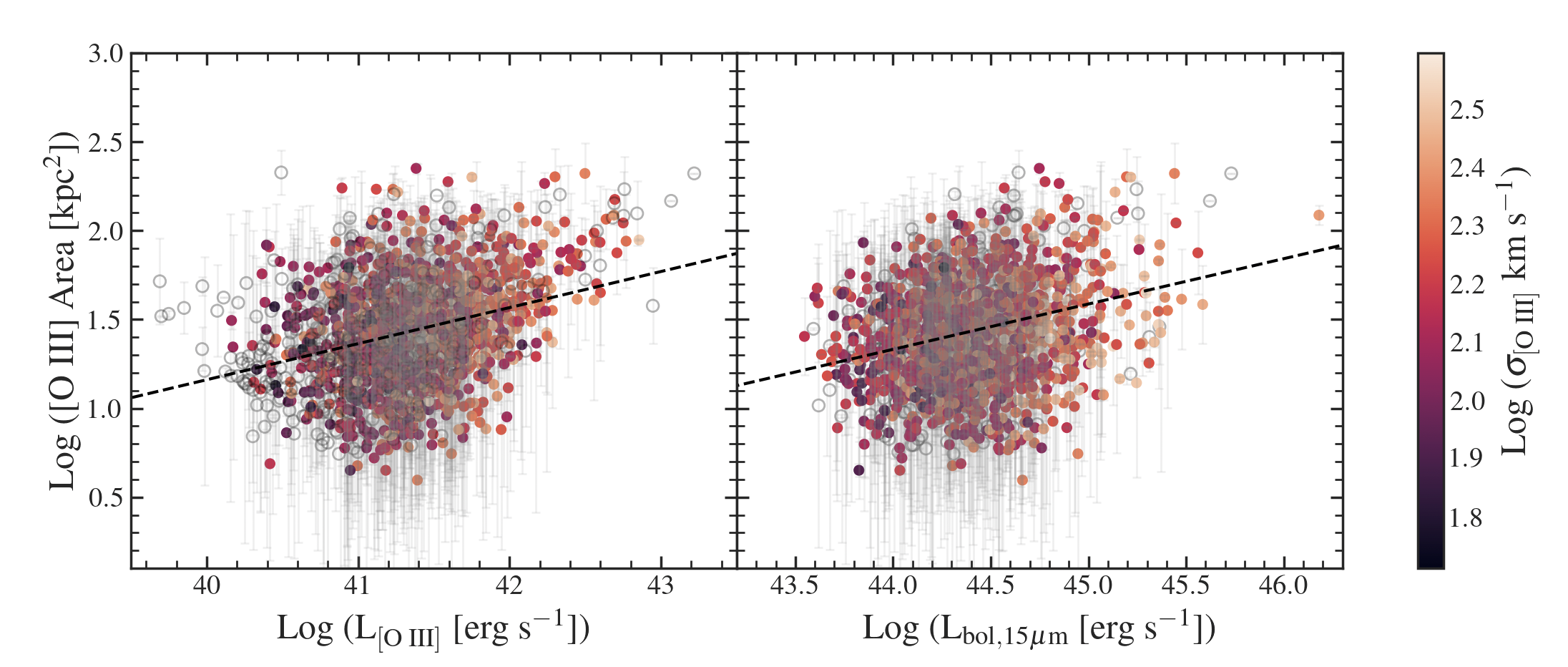}
\caption{
The \ot\ area as a function of \ot\ luminosity ($L_\mathrm{[O III]}$); left) and bolometric luminosity ($L_\mathrm{bol, 15\mu m}$; right). Solid circles are $2{,}009$ (left) and $1{,}945$ (right) AGNs, while their colors indicate the velocity dispersion of \ot\ emission (\secondm). Gray open circles represent objects with low-quality (Section \ref{sec:oiii_kin}). The dashed lines show the best-fit power laws from a linear regression model. The error bars indicate the standard deviation ($\sigma_\mathrm{A}$) in the area measurement (See Section \ref{sec:oiiiarea}).}
\label{fig:lum-areaall}
\end{figure*}

Figure \ref{fig:lum-areaall} illustrates the relationship between the \ot\ areas and luminosities with the velocity dispersion in the colorbar. 
Objects with higher luminosities tend to show larger \ot\ area and higher velocity dispersions. 
In the left panel, we plot the relation between \ot\ area and \ot\ luminosity without separating the samples at $z\geq0.26$ as plotted in Figure \ref{fig:areadist}.  
Based on Pearson's correlation coefficient, the \ot\ area is moderately correlated with the \ot\ luminosity ($r\sim0.33$, $p<0.001$). The correlation can be written as $\log (\mathrm{\ot\ Area}) = (0.20\pm0.01)\log (L_\mathrm{\ot})-7.02$. 
The right panel shows the plot between the \ot\ area and the bolometric luminosity computed from the rest-frame $15\mu m$ luminosity of $1{,}945$ subsamples with the WISE detections (see Section \ref{sec:wise}). 
The best-fit correlation is $\log (\mathrm{\ot\ Area}) = (0.27\pm0.02)\log(L_\mathrm{bol,15\mu m})-10.46$, with the Pearson's coefficient of $r\sim0.31$ at $p<0.001$.

We compare our results with the study by \citet{Sun_2018} using HSC broadband images to map the \ot\ images in $300$ type-2 AGNs. 
They have established the \ot\ Area - luminosity relation with a slope of $0.62\pm0.01$ and the Pearson's $r\sim0.68$ ($p<0.01$). 
Their correlation is significantly stronger than those found in this work. 
However, it is possibly due to the difference in the threshold used to determine the isophotal area.
\cite{Sun_2018} adopted a rest-frame isophotal threshold of $3\times10^{-15} $\sbunit, whereas we used the threshold of $1.4\times10^{-15} $\sbunit (Section \ref{sec:oiiiarea}).
They also showed the result of using the lower isophotal cut of $1\times10^{-15} $\sbunit, yielding a much flatter slope of $0.2$, which is comparable to our result. 
In addition, we noticed that they have a higher fraction of high-luminosity samples with \lbol$>10^{45}$ erg s$^{-1}$ that could affect the best-fit slope in a regression analysis. 
Nevertheless, our goal in this study is to justify the impact of outflows on the size-luminosity relation, and even so, our result could statistically verify that the size of the \ot\ emission is typically large among high-luminosity AGNs, which agrees well with previous studies \citep{Liu2013, Hainline2013, Sun_2017}. We make further discussions on the size-luminosity relation in Section \ref{discuss:size-lum}.

\subsection{[O\;III] spectral profiles and outflow signature}\label{sec:oiii_kin}

\begin{figure*}[ht!]
\epsscale{0.4}
\centering
\includegraphics[width=0.95\textwidth]{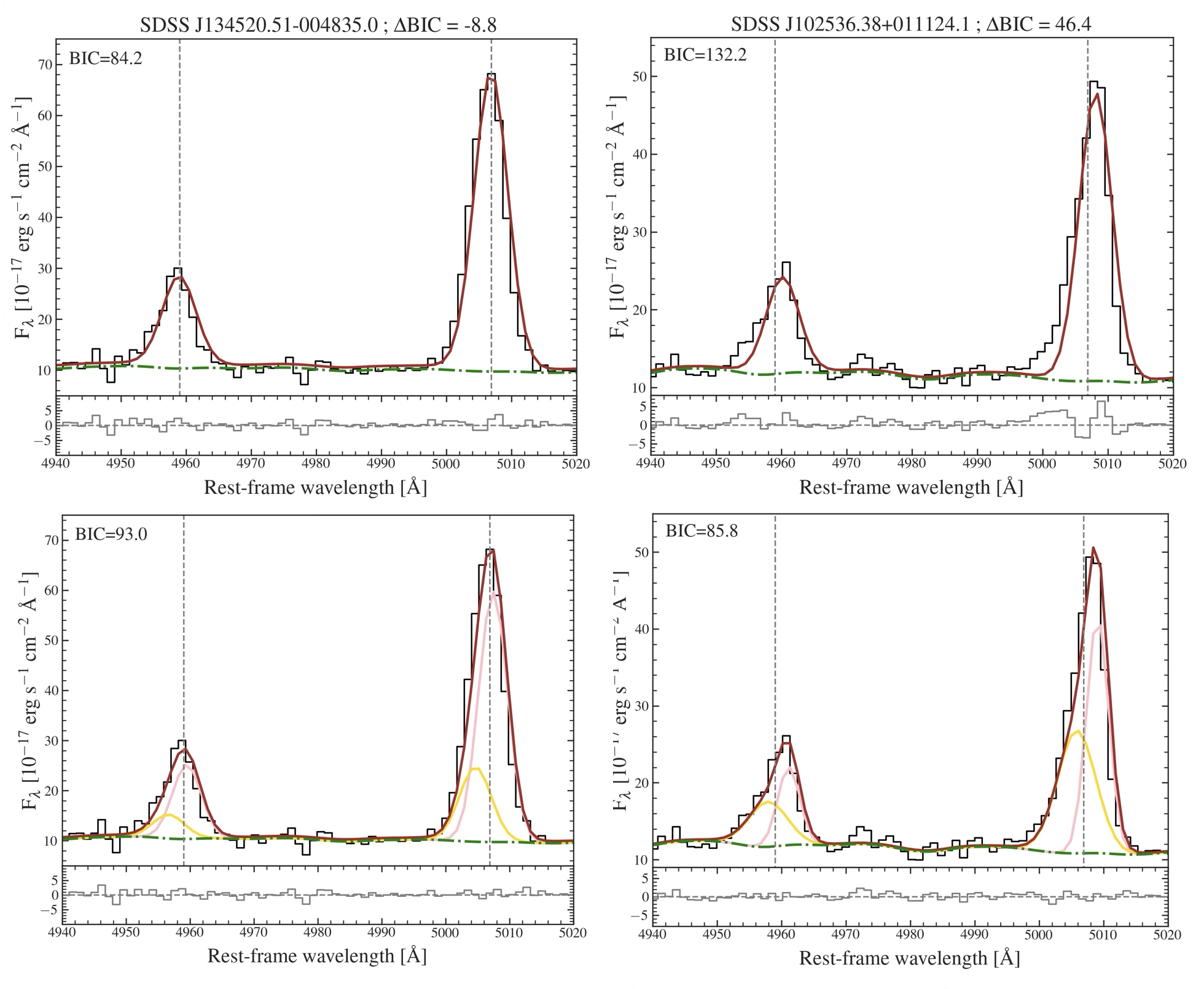}
\caption{\oiii\ emission lines  with the fitting residual of SDSS J$134520.51-004835.0$ at $z=0.17$ (left) and SDSS J$102536.38+011124.1$ at $z=0.18$ (right). The black solid line shows the original SDSS spectrum. The top and bottom panels show the case of using single and double Gaussian models, respectively. 
The red solid and green dash-dotted lines show the best-fit \ot\ emission lines and the stellar continuum model, respectively. The pink and yellow lines in the bottom panel represent the core and the wing components, respectively. The BIC value is also shown at the top left corner of each panel. The residual plots are displayed below each subplot.
\label{fig:emfitting}}
\end{figure*}

Two examples of \ot\ emission-line fittings are shown in Figure \ref{fig:emfitting}. The spectrum of each object is fitted with both single and double Gaussian models. The best-fit model is then determined from the $\Delta\mathrm{BIC}$ criterion, as explained in Section \ref{sec:oiiikinematics}. 
Basically, the model with the lowest BIC value is preferred. 
In the left panel, the \ot\ emission line of SDSS J$134520.51-004835.0$ at $z=0.17$ is better fitted with a single Gaussian profile, which has a lower BIC than the double Gaussian model. The $\Delta\mathrm{BIC}$ in this case is smaller than 10 ($\Delta\mathrm{BIC}=-8.8$). 
On the other hand, the double Gaussian profile is chosen in the case of SDSS J$102536.38+011124.1$ (right panel).
The double Gaussian model better describes the \ot\ profile, with $\Delta\mathrm{BIC}>10$. 
The wing component is relatively strong compared to the core component, with a wing-to-core flux ratio of $0.90$, making the total line profile look asymmetrical.
With this fitting procedure, we obtained $916$ and $629$ objects with single and double Gaussian profiles, respectively. 
The remaining $464$ objects were excluded due to their low spectral quality, i.e., more than $80\%$ of the flux density have the S/N$\leq8$ in the wavelength range of the \ot\ fitting.

\begin{figure}[ht!]
\epsscale{1.2}
\plotone{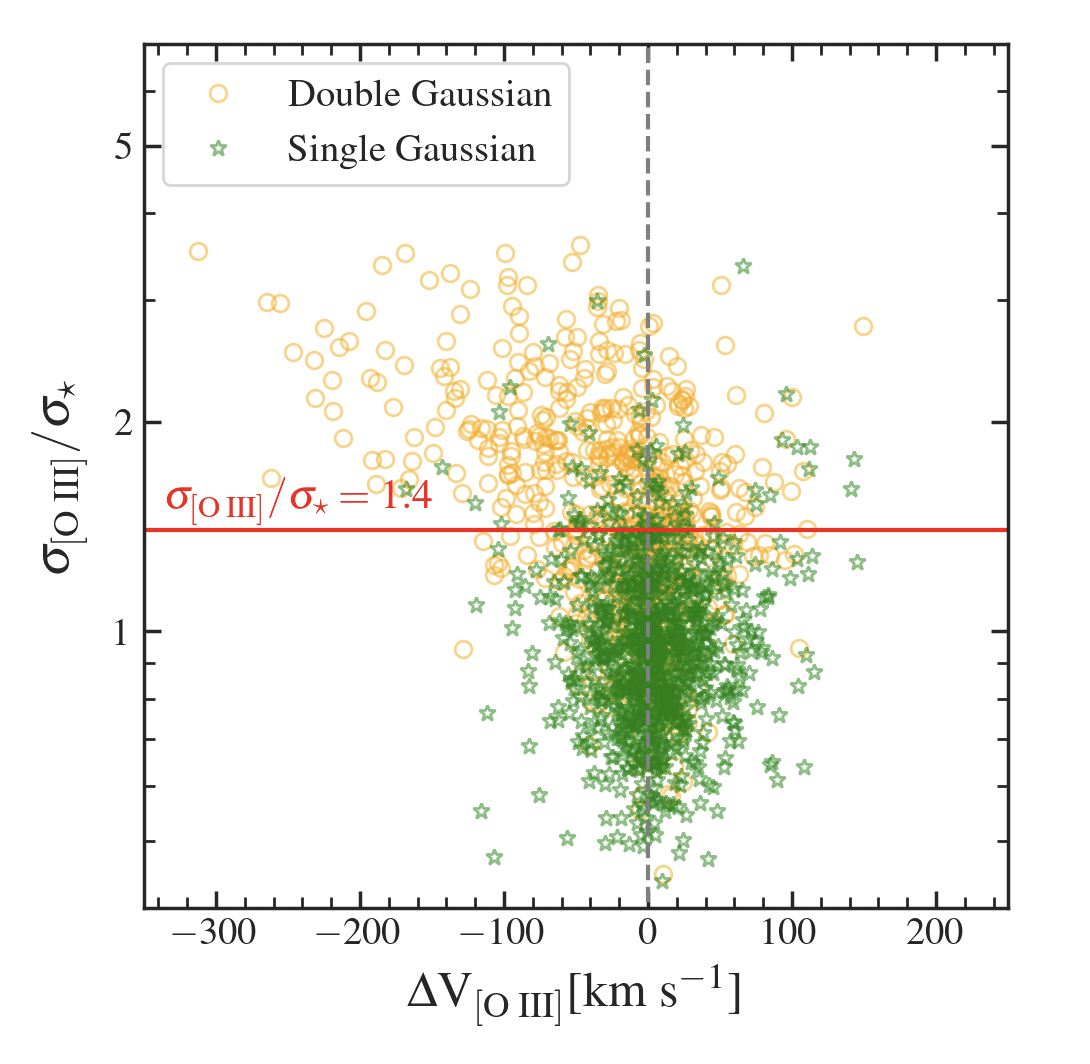}
\caption{The ratio of the \ot-to-stellar velocity dispersion versus velocity shift of $1{,}545$ objects. Yellow circles and green stars mark samples with the double-Gaussian and single-Gaussian profiles in their \ot\ emission lines, respectively. The red line marks the value of \secondm/\firstm$=1.4$ used to select sample with outflows (Section \ref{sec:oiii_kin}).}\label{fig:vel-vel.png}
\end{figure}

Several methods are used to identify outflow signatures in the galaxy based on the \ot\ emission-line profile, one of which is 
the presence of the wing component \citep[e.g.,][]{Wylezalek2018,Perna_2019,Zheng_2023}. 
However, the spectral fitting is sensitive to the spectral data quality. 
We may miss gas outflowing samples whose \ot\ emission line is best fitted with the single-Gaussian \ot\ profile due to low S/N ratios. 
In this study, we assume 
that the observed velocity dispersion of the \ot\ emission line is caused by both gravitational potential from the host galaxy and the non-gravitational component, which is interpreted as the outflow signature. That is 
\begin{equation}
        \sigma_\mathrm{[O\;III]}^2 = \sigma^2_\mathrm{gr} + \sigma^2_\mathrm{non-gr},
\end{equation}
where $\sigma_\mathrm{[O\;III]}$ represents the observed \ot\ velocity dispersion calculated from the best fitting Gaussian profile using Eq.\ref{eq:sigma}. 
The $\sigma_\mathrm{gr}$ and $\sigma_\mathrm{non-gr}$ denote the velocity dispersion of the gravitational and non-gravitational components, respectively. 
We adopted the stellar velocity dispersion to trace the strength of the gravitational potential, that is $\sigma_{\rm gr}=\sigma_\star$. 
We classified the samples with $\sigma_\mathrm{\ot}/\sigma_\star > 1.4$ as having the outflow signature. 
The $\sigma_\mathrm{\ot}/\sigma_\star = 1.4$ threshold is where the non-gravitational and gravitational components contribute equally to the observed \ot\ emission. 

Figure \ref{fig:vel-vel.png} shows the \secondm$/\sigma_\star$ ratio versus velocity shift (\firstm) of the samples. 
Generally, the \secondm$/\sigma_\star$ ratio tends to increase toward the negative \ot\ velocity shift. 
The samples with outflow signature selected by our $\sigma_\mathrm{\ot}/\sigma_\star > 1.4$ threshold are not only 
those whose \ot\ emission lines were best fitted with the double Gaussian profile, but also 
the samples with the best-fit single Gaussian \ot\ emission or with the low \ot\ velocity shift. 
For the galaxies with low dust extinction, it is possible that the receding component of biconical outflow can be detected and causes the lower estimation of the \ot\ velocity shift \citep{Woo2016}. 
As a result, we finally have $621$ and $924$ samples with and without the signature of the outflows, respectively. 

\subsection{The impact of outflows on the spatial extent of the [O\;III] region}\label{sec:result_kin}

\begin{figure*}[ht!]
    \centering   \includegraphics[width=0.9\linewidth]{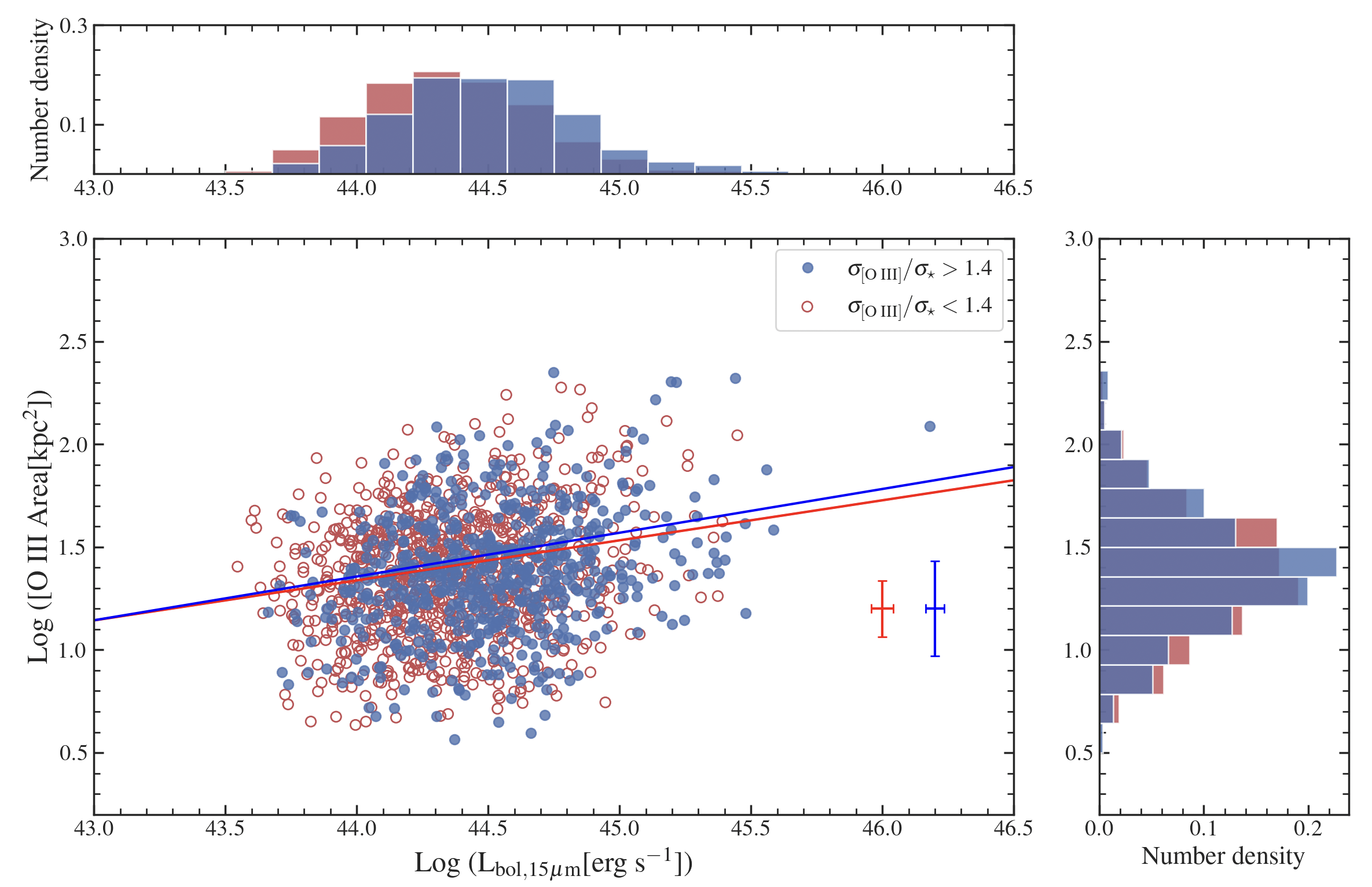}
    \caption{The \ot\ area-\lbol\ relations of AGNs with outflows (solid blue circles) and without outflows (open red circles). The solid blue and red lines show the best-fit relation of samples with and without outflows, respectively. The histograms along an x-axis and y-axis show the distributions of \lbol\ and \ot\ areas between two subsamples. The error bars denote the median $1\sigma$ uncertainty of the \ot\ areas and bolometric luminosity of two subsamples.}
    \label{fig:Lbol-area-kin}
\end{figure*}

In this section, we investigate whether the size of the \ot\ emission is influenced by the presence of gas outflows. We revisit the area-bolometric luminosity, as illustrated in Figure \ref{fig:Lbol-area-kin}, now separating the sample into objects with (blue) and without (red) outflows. 
The objects with outflows tend to have higher bolometric luminosity compared to those without outflows. 
Using the KS test, we can reject the null hypothesis that the bolometric luminosities of two subsamples are drawn from the same distribution ($p<0.01$). 
This suggests that outflows are more prominent among samples with high-luminosity AGNs.
In contrast, \ot\ area distributions of two subsamples are not statistically different. 
The best-fit slopes of the area–luminosity relation are $0.21 \pm 0.03$ for objects with outflows and $0.19 \pm 0.03$ for those without outflows
The correlation between \ot\ areas and the \ot\ velocity dispersion is also very weak, with Pearson's $r\sim0.12$ and $p<0.001$.
These results suggest that the presence of outflows does not significantly affect the extension of the \ot\ emission.
As the correlation of the area-luminosity relation is much stronger than the area-velocity relation, this suggests that the photoionization by central AGNs seems to be the main driver of the spatial extent of the \ot\ emission rather than the gas outflows.

\section{DISCUSSION}\label{sec:discuss}

\subsection{Detection of [O\;III] emission using broadband images}

The \ot\ images in this work are constructed based on the flux excess in the SDSS $r$-band images. 
Using the publicly available broadband data is an alternative inexpensive approach to increase the sample size in the study of the NLR. 
Previous studies have found a number of compact galaxies with strong \ot\ emission lines giving distinct flux excess in the SDSS filters \citep{Cardamone2009, Liu_2022}. 
With $\sim 1370$\AA\ FWHM of the $r$-band filter, the \ot\ emission line with the observed equivalent width EW(\ot) $\sim100$\AA\; contributes to a magnitude excess of $\Delta m_\mathrm{AB}\sim0.1$ in the emission-line image. 
Only galaxies with strong \ot\ emission would be detected with this method. 

\begin{figure}[ht!]
\epsscale{1.20}
\plotone{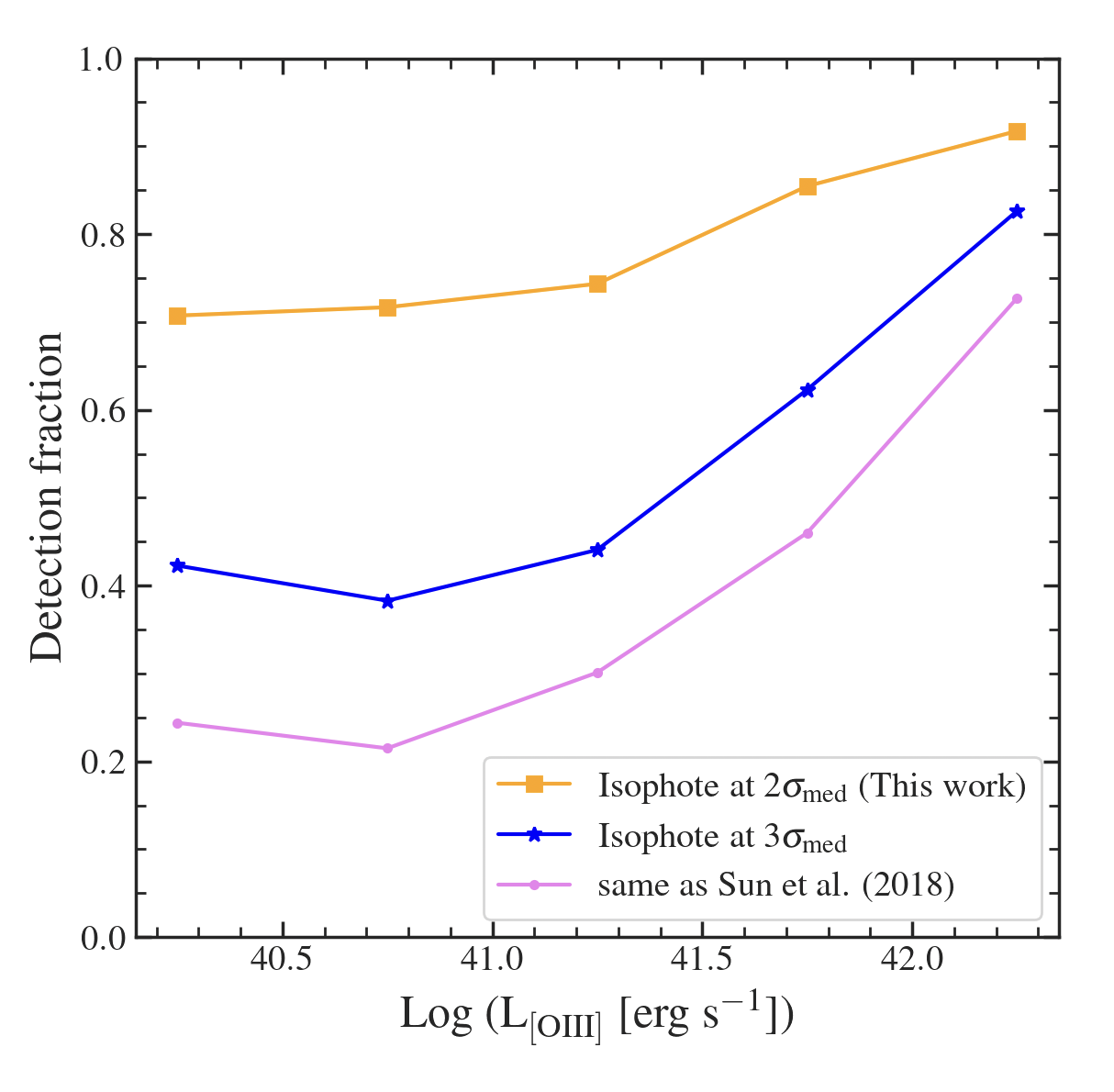}
\caption{Detection fraction of \ot\ emission using our broadband selection technique as a function of \ot\ luminosity. Orange squares, blue stars, and magenta dots represent the rest-frame isophotal thresholds at $2\sigma_\mathrm{med}$, $3\sigma_\mathrm{med}$, and $3\times10^{-15}$ \sbunit, respectively.}\label{fig:detection-fraction.png}
\end{figure}

As the value of EW(\ot) typically increases with the \ot\ luminosity \citep{zakamska_2003}, it is expected that the number of objects with detected \ot\ emission should increase with \ot\ luminosity. 
The fraction of detecting \ot\ emission in the broadband excess images or emission-line image is plotted as a function of \ot\ luminosity in Figure \ref{fig:detection-fraction.png}. 
We compared three different isophotal criteria used in the area measurement: $
1.4\times10^{-15}$ \sbunit\ $(2\sigma_\mathrm{med})$,\ $2.1\times10^{-15}$  \sbunit\ $(3\sigma_\mathrm{med})$, and $3\times10^{-15}$ \sbunit\ (used in \citealt{Sun_2018}). 
It is clearly seen from the figure that the detection fraction increases with the \ot\ luminosity in all cases. 
However, the fraction reduces greatly when using the brighter isophotal threshold. 

\subsection{Comparison of size-luminosity relation}\label{discuss:size-lum}

We found that the \ot\ area constructed from our broadband technique is well correlated with both \ot\ and AGN luminosity consistent with previous studies \citep{Hainline2013, Sun_2017, Sun_2018, Polack_2024}.
Although many studies found a strong correlation between the radial extent of the NLR and AGN luminosity in type-2 AGNs, 
the power-law slopes in the size-luminosity relation are not consistent with each other \citep{Liu2013,Hainline2013}.
To compare our results with these studies, we calculated the maximum extent from the center of [O III] emission to the farthest point within the isophote.
We found a very weak correlation of the maximum extent with the bolometric luminosity (Pearson's $r\sim 0.15$) with a slope of $0.11\pm0.02$, which is much lower than the aforementioned studies.
We found that the measurement of the maximum extent is not accurate for broadband imaging. The maximum extent is measured from the center to the maximum distance where the intensity is just above the isophotal threshold.
Consequently, we used the isophotal area to represent the spatial extent of the [O III] emission.

The study using the Subaru/HSC broadband images by \citet{Sun_2018} shows a strong correlation between the \ot\ area and AGN luminosity with the power-law slope of $0.62\pm0.10$. They measured the \ot\ areas down to the rest-frame isophote of $3\times10^{-15}$ \sbunit.
\citet{Hainline2013} and \citet{Sun_2017} reported the shallower power-law slopes of $0.47\pm0.02$ and $0.30$, respectively, when a slightly deeper surface brightness cutoff was adopted at $10^{-15}$\sbunit. 
Using the narrowband HST images, \citet{Polack_2024} adopted a very faint isophotal threshold of $2\times10^{-17}$\sbunit\ and found a power-law slope of $0.39\pm0.05$.

The discrepancy in the derived power-law slopes is possibly due to different surface brightness thresholds used to determine the size of the \ot\ emission. Using a brighter isophote would likely increase the slope of the size-luminosity relation.

\begin{figure}[ht!]
\centering
\includegraphics[width=1.03\linewidth]{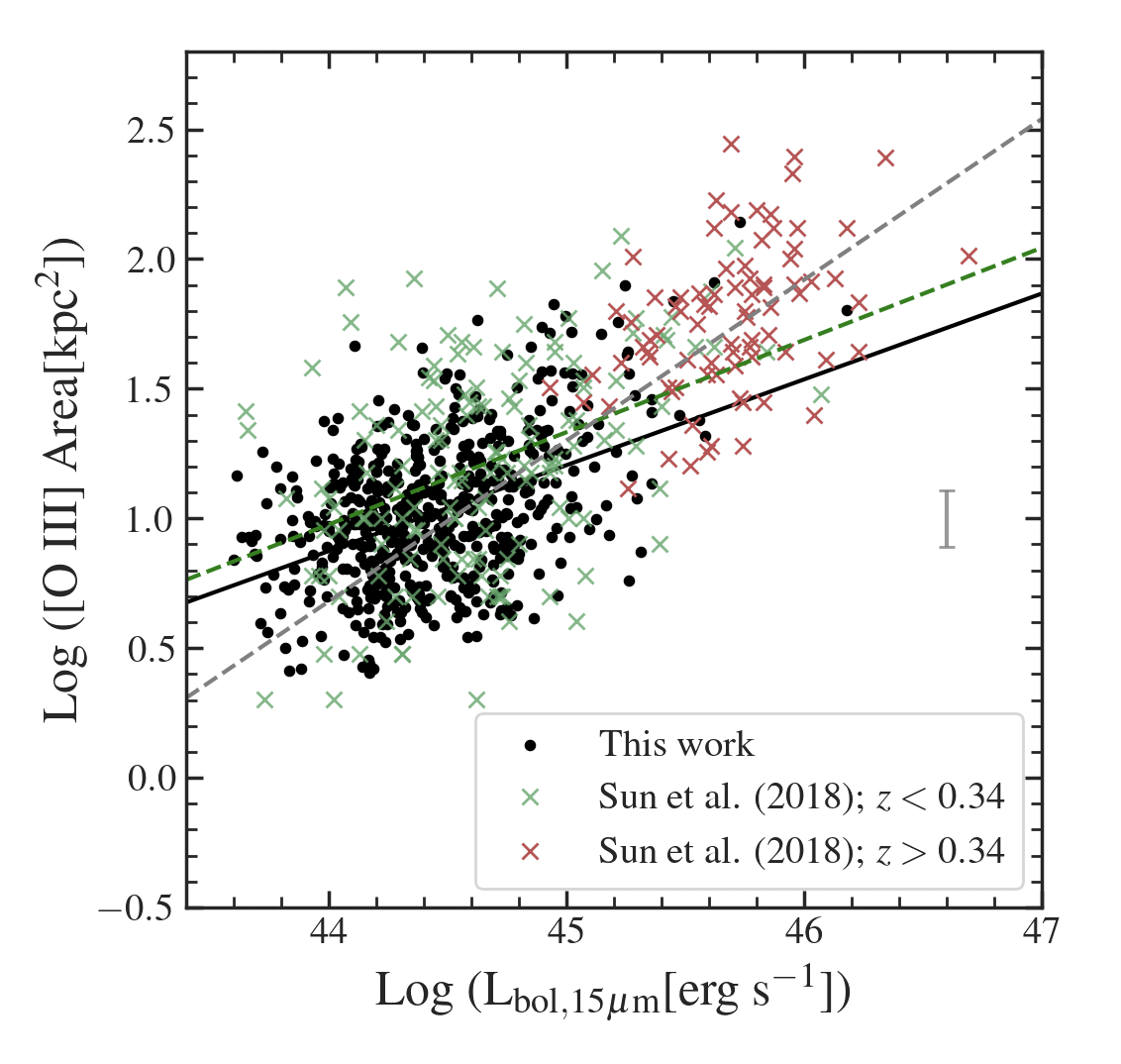}
\caption{The \ot\ isophotal area - bolometric luminosity relation. Black circles are samples from this work, while green and red cross symbols represent objects from \citet{Sun_2018} at $z<0.34$  and $z>0.34$, respectively. The black solid line shows the size-luminosity relation for our samples. The gray dash line indicates the original size-luminosity relation from \citet{Sun_2018}, while the green dash line corresponds to the relation of their samples at $z<0.34$.}
 \label{fig:Lbol-area-sun}
\end{figure}

To confirm this hypothesis, we compared our results with the study by \citet{Sun_2018}, who used the same broadband excess technique but with the Subaru/HSC images and significantly brighter 
isophotal threshold ($3\times10^{-15}$ \sbunit). 
To make a fair comparison, we adopted the same isophotal threshold and remeasured the \ot\ area of our samples. 
The new \ot\ area - \lbol\ relation is plotted in Figure \ref{fig:Lbol-area-sun} along with $300$ type-2 AGNs from \citet{Sun_2018}. 
Because \citet{Sun_2018} included the samples at higher redshift than our work, we separated the samples from \citet{Sun_2018} into two redshift ranges ($z<0.34$ and $z>0.34$). The figure shows that our samples have the [O III] isophotal areas and bolometric luminosities of the two samples consistent with those of \citet{Sun_2018}, when considering only objects at $z<0.34$ (green). Their higher-redshift sources ($z > 0.34$) are predominantly more luminous and exhibit larger \ot\ areas.
The \ot\ areas of our samples are less scattered at the low end of luminosity, while we have a significantly smaller number of samples at higher luminosity. 
With this new isophotal threshold for the \ot\ area measurement, the correlation between the \ot\ areas and luminosity of our samples becomes stronger with the Pearson's $r\sim0.52$, $p<0.001$ (cf. Section \ref{sec:area-lum}). 
This correlation strength becomes closer to $r\sim 0.68$ reported in \citet{Sun_2018}. 
We obtained a slightly steeper slope of $0.39\pm0.02$ than using a shallower isophotal threshold in Section \ref{sec:area-lum}, but it is still shallower than the study by \citet{Sun_2018}. 
The difference in the slope might be due to the lack of high-luminosity samples (\lbol$>10^{45}$ erg s$^{-1}$), which could affect the result of the best-fit slope. 

\subsection{Influence of outflows on the gas clouds in the narrow-line regions (NLRs)}\label{sec:area-kin}

It is known that the ionization of gas in the NLR is mostly from the power-law radiation from the central AGNs \citep{Evans&Dopita_1985, Binette1996}, sometimes with the contribution of fast shocks \citep{Terao_2016}.
The formation of the gas clouds in the NLRs is still unclear. One explanation is that they are already located inside the host galaxy and is photoionized by central AGNs.
On the other hand, based on the radiation-driven fountain model by \citet{Wada_2018}, they showed that the AGN-driven outflow could also transfer the gas cloud from the central region to the outer part. Although the simulation is spatially-limited within $\sim16$ pc from the nucleus, the more extended NLRs could be inferred from the model.
The observational study by \citet{Joh2021} also supports the scenario that the gas clouds in the NLRs are originated in the nucleus, which are transferred by gas outflows as those gas have higher velocity dispersion and electron density compared to gas in the \textsc{Hii} region.

Our assumption is that if the gas in the NLRs is driven by outflows, their \ot\ areas should be statistically different compared to samples without outflows, or at least the slope in the size-luminosity relation should be different.
However, based on our results in Section \ref{sec:result_kin}, it is found that two subsamples show very similar slopes in the \ot\ Area-\lbol\ relations (Figure \ref{fig:Lbol-area-kin}). The distribution of \ot\ areas between two subsamples is also statistically consistent, with a $p$-value of $\sim0.17$ in the KS test and $p>0.25$ in the AD-test.
It seems that the presence of gas outflows has no effect on the size of the \ot\ emission.

In addition, we investigate the correlation of the area with the \secondm/$\sigma_\star$ ratio, which represents the strength of outflows. 
We have found that the ratios have much weaker correlation with the \ot\ areas ($r\sim0.1$) compared to the area-luminosity relation ($r\sim0.3$). 
It is unlikely that the non-gravitational kinematics, or outflows alone could create more extended \ot\ emission.

Our results do agree with previous spectroscopic studies, who found that the size of outflows is usually smaller than the size of the NLRs with the ratio of $0.22-0.72$ \citep{Fischer_2018, Kim_2023, Polack_2024}. 
On the other hand, the recent study by \citet{Zheng_2023} found that the size of gas outflows in the dwarf galaxy SDSS J$0228-0901$ is significantly larger than the NLR size, while the latter is still not deviated much from the standard slope of $\sim0.42$ in the size-luminosity relation by \citet{Chen_2019}.
These findings suggest that, in most cases, the outflow kinematics should not affect the size of the NLRs, in which the AGN photoionization is often dominant at larger distance.

\section{SUMMARY}\label{sec:conclusion}

This work investigates the impacts of gas outflows on the size of the NLRs, which is traced by the region of \ot\ emission.
We constructed the \ot\ maps using the SDSS broadband images in the $r$-band filter, together with a continuum subtraction interpolated from the $i$ or $z$ band using the SDSS spectra.
Our broadband technique can significantly enhance the sample size in the study of the NLRs compared to other spatially resolved techniques.
Using a sample of type-2 AGNs at $0.13<z<0.34$ from the SDSS ``emissionlinesport" table, there are $2{,}009$ objects with the detection of \ot\ emission.
Among all samples, there are $1{,}545$ objects with reliable \ot\ emission line to determining the presence of outflows.
The main results are summarized as follows.
\begin{enumerate}
\setlength\itemsep{-.1em}
    \item{Based on the rest-frame isophotal cut of $1.4\times10^{-15}$ \sbunit, our samples have the \ot\ areas ranging from $3.7$ kpc$^2$ up to $224$ kpc$^2$. The \ot\ areas seem to increase with the \ot\ luminosity.}
    \item{With the AGN bolometric luminosity calculated from the rest-frame $15\mu$m WISE luminosity, we obtain strong correlation of the \ot\ area with both \ot\ and AGN bolometric luminosity with the Pearson's $r=0.33$ and $r=0.31$, respectively. 
    We have established the area-luminosity relation as $\log (\mathrm{\ot\ Area}) = (0.27\pm0.02)\log(L_\mathrm{bol,15\mu m})-10.46$.}
    \item{By using the same isophotal level by \citet{Sun_2018} in the area measurement, the \ot\ areas of our samples are consistent with their values. The correlation of the area-luminosity relation becomes stronger with Pearson's $r$ of $0.52$ which is close to their study.
    However, the slope is still lower compared to their study}. We believed that it is caused by the lack of high-luminosity AGNs in our samples, which could affect the slope in the regression analysis.
     
    \item{As the \ot\ luminosity increases, the \ot\ velocity dispersion is higher and the velocity shift becomes more negative, suggesting the outflow kinematics is stronger in more luminous AGNs.}
    \item{Using the strength of the non-gravitational component $\sigma_\mathrm{[O\;III]}/\sigma_\star$ ratio at $1.4$, we are able to classifiy $621$ and $924$ objects as having outflows and no outflows, respectively.}
    \item{The samples with and without outflows share very similar slopes in the area-luminosity relation.} The area of the NLR in AGNs with outflows is not statistically different from that in AGNs without outflows.
\end{enumerate}

Our findings suggested that the extended \ot\ emission is not influenced by outflow mechanisms, questioning the effectiveness of the mechanical mode of the AGN feedback to deplete the gas reservoir from the host galaxies particularly in the moderate luminosity regime ($10^{43}-10^{46}$ erg s$^{-1}$). 

\section{Acknowledgements}
We would like to thank the anonymous reviewer for helpful comments and suggestions that improved the clarity of this paper. This research project is supported by the National Research Council of Thailand (NRCT): N41A640219.
S.Y. is supported by the Office of the Ministry of Higher Education, Science, Research, and Innovation through a research grant for new scholars (RGNS63-175).
\vspace{5mm}

\bibliography{Reference}{}

\end{document}